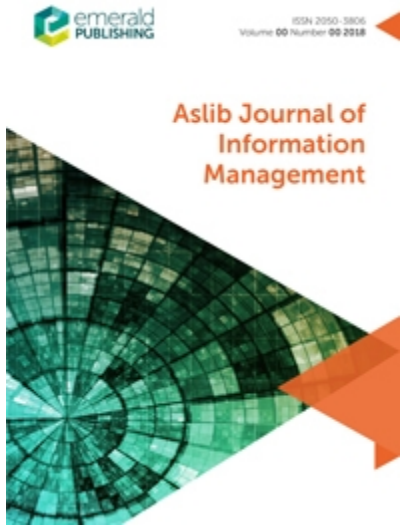

# Public interest in science or bots? Selective amplification of scientific articles on Twitter

# Public interest in science or bots? Selective amplification of scientific articles on Twitter

Ashiqur Rahman[1], Ehsan Mohammadi[2], and Hamed Alhoori[3]

## Abstract

**Purpose**

With the remarkable capability to reach the public instantly, social media has become integral in sharing scholarly articles to measure public response. Since spamming by bots on social media can steer the conversation and present a false public interest in given research, affecting policies impacting the public's lives in the real world, this topic warrants critical study and attention.

**Methodology**

We used the Altmetric dataset in combination with data collected through Twitter API and the Botometer API. We combined the data into a comprehensive dataset with academic articles, several features from the article, and a label indicating whether the article had excessive bot activity on Twitter or not. We analyzed the data to see the possibility of bot activity based on different characteristics of the article. We also trained machine-learning models using this dataset to identify possible bot activity in any given article.

**Findings**

Our machine-learning models were capable of identifying possible bot activity in any academic article with an accuracy of 0.70. We also found that *Health and Human Science* related articles are more prone to bot activity compared to other research areas. Without arguing the maliciousness of the bot activity, our work

---

[1] Department of Computer Science, Northern Illinois University, DeKalb, Illinois, USA

[2] College of Information and Communications, University of South Carolina, Columbia, South Carolina, USA

[3] Department of Computer Science, Northern Illinois University, DeKalb, Illinois, USA

presents a tool to identify the presence of bot activity in the dissemination of an academic article and creates a baseline for future research in this direction.

## Originality

While the majority of the existing research focuses on identifying and preventing bot activity on social media, our work is novel in predicting the possibility of bot activity in the dissemination of an academic article using Altmetric metadata for the article. Little work has been performed in this specific area, and the models developed from our research give the policymakers and the public a tool to interpret and accept the public interest in a scientific publication with appropriate caution.

## Keywords

## Citation

# Author Biography

**Ashiqur Rahman**

(ORCID: 0000-0003-3290-2474)

Ashiqur is a PhD student in the Department of Computer Science at Northern Illinois University, DeKalb, Illinois, USA. His current works involve data analysis, developing machine-learning models, NLP, and computer vision.

**Ehsan Mohammadi**

(ORCID: 0000-0003-3481-6991)

Ehsan is an Associate Professor in the School of Information Science at the University of South Carolina. He uses data science and web analytics methods to answer questions in different contexts, such as health informatics, mis/disinformation, social media analysis, and the application of AI.

**Hamed Alhoori**

(ORCID: 0000-0002-4733-6586)

Hamed is an Associate Professor (tenured) in the Department of Computer Science at Northern Illinois University and is the Director of the Data Analytics Theory and Applications (DATA) Laboratory. His research aims to advance new data-driven scientific discoveries by quantifying dynamic global research patterns and needs, mining, and learning from massive unstructured scholarly datasets, modeling emerging multidimensional web indicators, recommending scholarly content pertinent to researchers' activities and interests, and predicting the societal impact of research.

# 1 Introduction

Dissemination of academic articles through social media can be a valuable metric to determine their impact on the public interest (Bornmann, 2015). Altmetric gives us precisely that as a numeric score for any academic article. Altmetric is an alternative to the traditional citation-based metrics, which includes peer reviews on Faculty of 1000, citations on Wikipedia, discussions on research blogs, mainstream media coverage, bookmarks on reference managers like Mendeley, and mentions on social networks such as Twitter (Altmetrics, 2015; Erdt et al., 2016). Because of its high reliance on social media activity, the Atlmetric score is likely to be vulnerable to bot activities on social media.

The study by Torres-Salinas et al. (2018) suggests that Altmetric does not provide a complete picture of the social attention of a book and is rather vulnerable to automated bots. In a similar context, our current study only looks at the Twitter dataset from Altmetric and the activity of Twitter bots in that dataset. The term "bot" here refers to automated social media accounts that act with specific goals and often prevent the organic growth of a conversation (Ferrara et al., 2016; Liu, 2019; Albadi et al., 2019). Besides the Twitter user information and Altmetric score, the dataset contains several important features for each article, like the paper's origin, publication time, research area, and where it was published.

Being one of the most prominent social media globally, and with the ability to quickly spread information to a larger population (Ozturkcan et al., 2017), Twitter has become a preferred platform for scholarly communication. A higher tweet count can indicate the higher social importance of a publication and influence the news mentions (Htoo et al., 2023), commencing higher dissemination. However, Twitter is also prone to bot and spam activity. "Spambot" accounts on this platform are especially concerning because they can quickly generate vast traffic and influence people's opinions (Efthimion et al., 2018). Much research on Twitter data focuses on identifying bots (Liu, 2019; Chavoshi et al., 2016a,b; Lokot and Diakopoulos, 2016; Minnich et al., 2017; Zeng et al., 2021), and there is a growing concern that bot activity may significantly change the public opinion on crucial scientific discussion (Broniatowski et al., 2018; Jamison et al., 2020; Didegah et al., 2018; Ferrara, 2020). Following the observation of Burstein (2003) regarding the substantial influence of public opinion on policy decisions, it becomes crucial to investigate the impact of social bots on the dissemination of scientific articles and their subsequent reception within the public domain.

With the increasing influence of social media on everyday life, many research areas and researchers may intentionally or unintentionally become a target of social media bots (Didegah et al., 2018; Estella and Na, 2018; Jamison et al., 2019; Ledford, 2020). We found that bot activity within academic tweets is prevalent across many countries, with a particularly notable rise in activity observed in some developing countries. This bot activity can prevent the organic growth of any conversation and can present a false public interest in a given scientific article, leading to policies impacting people's lives. Even though not all bots are harmful, malicious bots can take advantage of this situation. Therefore, in our research, we use features of any article and aim to predict the possibility of high bot activity using different machine learning models.

Although Twitter bot activity is a well-researched topic, our research is directed to a novel approach to building a machine learning model to predict the possibility of whether there will be higher bot activity for any given research article. Our analysis also identified that health and human science articles are more prone to higher bot activity than other research areas. Recognizing the various functionalities of social bots (Lokot and Diakopoulos, 2016; Oentaryo et al., 2016), our study focuses on identifying the possible presence of bot activity within an academic tweet discussion without judging the presence as positive or negative. This model can be helpful for researchers, decision-makers, and the general population as a whole.

We also acknowledge that the platform "Twitter" has been renamed as "X," and the "tweets" are now called "posts" in the new platform (Davis, 2023). However, to maintain readability, avoid potential confusion for readers unfamiliar with the change, and leverage the established name recognition, we will refer to the platform and the contents of the platform in its former name in this study.

## 1.1  Research Questions

In our research, we collected and analyzed the Altmetric dataset for Twitter to identify the presence of bot activity in academic publications. Throughout our analysis, we answered the following research questions:

1. Can we build a reliable metric to identify bot activity in disseminating academic publications?
2. Is there a significant difference in the prevalence of bot activity across various academic disciplines?

While answering these questions through this study, we have developed and published the following datasets and models. These resources are freely available to the research community to facilitate further research and advance the knowledge in this area. Links to the dataset and model are provided in Appendices A and B.

- A comprehensive dataset of scholarly articles with binary labels indicating higher activity of Twitter bots.
- Machine learning model to identify possible bot activity on an academic article.

## 2 Related Works

### 2.1 Bot Activity in Social Media

On the social media landscape, social bots are defined as automated accounts designed to mimic human behavior and generate content, with purposes ranging from information dissemination to gaining influence or steering the conversation. These bots can be categorized as informational or malicious based on their purpose (Oentaryo et al., 2016; Orabi et al., 2020). These social bots are a long-studied but unsolved topic in the research community (Cresci, 2020; Aljabri et al., 2023). The frequency of social bots and their activity significantly thwart different scientific advancements. Besides creating a nuisance and wasting valuable time, they can shift public opinion (Broniatowski et al., 2018) and have a real-world impact. We see that with the anti-vaccination (Broniatowski et al., 2018; Jamison et al., 2019) movement, spreading conspiracy theories (Jamison et al., 2020; Ferrara, 2020; Kitzie et al., 2018; Islam et al., 2020), and most recently with COVID-19 related misinformation dissemination (Jamison et al., 2020; Ferrara, 2020). Besides political and nefarious purposes, businesses also use bots to boost their sales and inflate their presence on social media (Cresci et al., 2014; Aswani et al., 2018; Obadimu et al., 2019).

Although not all bots are malicious (Lokot and Diakopoulos, 2016; Oentaryo et al., 2016), some can have specific agendas and cause severe social discord (Broniatowski et al., 2018; Kitzie et al., 2018; Stukal et al., 2019). Researchers have tried to understand the impact of bot activity and how it may steer social media conversation. After analyzing a large corpus of news links shared on Twitter, Shao et al. (2018) found that bots disproportionately share content from less credible sources. Mønsted et al. (2017) examined whether information adoption is done with simple exposure (single source) or complex exposure (multiple sources). The authors used social bots (39 coordinated bots) to spread information and determine the effect. They conclude that the complex contagion model spreads information more reliably. Liu (2019) examined 29 million tweets and found that social bots severely distort information and Twitter social bots are significantly more effective at spreading word of mouth.

The rebranding of Twitter to "X" introduced significant changes to their APIs, limiting free access to historical tweets (Ledford, 2023; Twitter, 2023). This presents a hurdle for academic research that relies on such data. However, the free API access still provides write permission, leaving the door open for bot activity. This underscores the ongoing need for research in effective bot detection methods.

## 2.2 Identifying Social Bots

Bot activity detection on social media has been a significant area of research interest. Although there is presence of bot activity, applications, and bot detection research on different social media platforms (Vukovic and Dujlovic, 2016; Akyon and Esat Kalfaoglu, 2019; Kim et al., 2021), for the purpose of this study, we will focus on Twitter bot activity.

Bot activity makes it difficult to build any actual metric and find reliable results by analyzing social media data to make an informed decision. Ongoing research proposes different methods to identify and clean the bot activity from social media data. Minnich (2017) proposed removing all noise from the text using NLP techniques and 'Behavior Profiling' and 'BotWalk' (Minnich et al., 2017) to remove bot activity from the data. For 'Behavior Profiling,' an extensive collection of features for any user is compiled and analyzed to determine bot-like behavior. Then, the 'BotWalk' algorithm uses the behavioral feature vector to identify potential bots in real-time.

Bots frequently change their approaches to social media, making it very difficult to detect their behavior (Sayyadiharikandeh et al., 2020). This adaptive behavior makes improving the bot-detection algorithms a continuous process. Several methods to detect bots and spammers have been proposed (Efthimion et al., 2018; Chavoshi et al., 2016a; Minnich et al., 2017; Aswani et al., 2018; Davis et al., 2016; Subrahmanian et al., 2016; Dhawan and Simran, 2018; InuwaDutse et al., 2018; Yang et al., 2019; Arin and Kutlu, 2023) by researchers over the years.

Chavoshi et al. (2016a) proposed a bot detection method, 'DeBot,' which uses warped correlation. The proposed method detects bots based on synchronized and correlated activity between users. Another approach Chavoshi et al. (2016b) proposed depends on correlated activities between Twitter accounts. They used Amazon Mechanical Turk to validate the detections. Minnich et al. (2017) proposed another near real-time bot detection method. Since modern bots keep adapting their behavior to evade detection, the proposed method - BotWalk, uses a model that evaluates different Twitter account features to identify possible bot accounts. Efthimion et al. (2018) analyzed the prevalence of bots and introduced a new algorithm to detect Twitter bots, achieving as low as 2.25% misclassification rate. Feng et al. (2021) proposed a comprehensive dataset to benchmark different bot detection methods and their adaptability with the evolution of bot behavior. Zeng et al. (2021) proposed a semisupervised model to identify bots on Twitter with improved performance and better at labeling fake accounts from imbalanced data. Davis et al. (2016) introduced a bot activity evaluation tool called 'BotOrNot,' which uses supervised machine-learning techniques to evaluate several features of Twitter user profiles and assigns a score for the possibility of a bot account. Several version updates of the tool have been released over the years, and it was renamed to 'Botometer' (Yang et al., 2022, 2023). The bot detection tool achieved promising results, with the latest version 4 achieving 0.99 area under the curve (AUC) performance, and offers a score between zero and five on different metrics, evaluating Twitter accounts as bots. The advent of artificial intelligence also made the bots more efficient in avoiding detection. Arin and Kutlu (2023) proposed a deep learning-based method utilizing long-short term memory (LSTM) to detect bot activities and achieved promising results.

Yang et al. (2019) reviewed different bot detection techniques and the use of artificial intelligence for this purpose. The authors used Botometer (Davis et al., 2016) to evaluate the public interaction with bot detection tools. Aljabri et al. (2023) also reviewed machine-learning-based bot detection techniques on different social media platforms, namely Facebook, Twitter, Instagram, LinkedIn, and Weibo. The authors found that Twitter is the most researched social media for bot activity and identified other potential areas for future research.

However, social bot identification is still an ongoing research without a universal method to identify them. Martini et al. (2021) analyzed three different bot detection techniques and concluded that different methods provide vastly different results, causing a reliability problem in the bot detection tools. The dependency and limitations of the API and the training data for the detection tool also dictate the performance of any given model (Yang et al., 2023).

Considering all the available tools, we chose to use Botometer (Davis et al., 2016; Yang et al., 2022, 2023; The Observatory on Social Media, 2016) to identify bots because this tool offers us detailed scores in different metrics and gives us the control to set the threshold to label a Twitter user as a bot. The tool had to evolve to adapt to the recent changes in Twitter API policy and was renamed "Botometer X" (Botometer, 2023). Although the latest version of the tools performs at a limited capability, these changes do not affect the findings of our study and the validity of our machine-learning models.

## 2.3     Scholarly Publications on Social Media

With the high popularity of social media, more research is being shared and discussed on these platforms (Liu et al., 2020). This phenomenon has sparked the interest of the research community in exploring the impact of social media mentions on scholarly articles and their potential as alternative metrics (Bornmann, 2015; Torres-Salinas et al., 2018; Priem and Costello, 2010; Costas et al., 2015a).

The rapid dissemination and instantaneous feedback made Twitter an effective platform for scholarly exchange. The study by Priem and Costello (2010) examined the behavior of academics tweeting scientific publications and found that citing on Twitter is often done as part of a conversation and expands beyond the domain expertise of the user. They found that academics prefer Twitter citations for faster dissemination and believe these citations represent scholarly impact. Schnitzler et al. (2016) also suggests that Twitter as a platform offers rapid, broad, and cost-effective opportunities for researchers to disseminate their work, which cannot be ignored. Moreover, the analysis of scholarly tweets from the American Journal of Neuroradiology (AJNR), Wadhwa et al. (2017) found a high engagement rate with tweets containing hashtags and images. These findings collectively illuminate the significance of Twitter as a tool for scholarly communication within academic communities.

However, there is a lack of consensus regarding the scholarly impact of Twitter engagement. Costas et al. (2015a) compared the social media mentions of scholarly articles from different disciplines and found differences among platforms and disciplines, suggesting a difference in social media preferences among researchers of different disciplines. The study by Bornmann (2015) suggests that social media mentions indicate the social impact of research. However, the popularity and frequency of shares on social media do not always translate into citations. In an earlier study, Eysenbach (2011) reported that tweets could predict highly cited articles within the first three days of the publication. But Didegah and Thelwall (2018) conclude that counts of tweets are not a reliable metric because many researchers may tweet or save articles from other researchers they follow or work with but are not quite interested in the area. The actual correlation between tweets and citations was very low. Bot activity also skewed the measurement. The experiment found that among all the Twitter users in their dataset, almost 46% of prolific article tweeters are bots, while 21% of moderate and 11% of occasional article tweeters are bots. Thelwall (2021) also listed the potential limitations of social media-based alternate metrics to measure scholarly impacts and advocated caution when using these metrics for research.

Estella and Na (2018) found that citation-oriented and Twitter mentionoriented academic articles have different characteristics, and high social media activity for an academic article does not guarantee the merit or quality of the work. In a study on manuscripts shared on Twitter, Carlson and Harris (2020) analyzed tweets of bioRxiv pre-prints and concluded that higher dissemination on social media does not ensure higher public exposure. They found that the majority of the audience on Twitter is from academia. Bowman (2015) found that there are statistically significant differences between the professional and personal tweets of scholars. In their study, Vainio and Holmberg (2017) reported that ideologically divisive research is shared more often, and the users sharing academic articles present themselves more professionally. The findings from Costas et al. (2015b) also suggest there is a weak correlation between Altmetric score and citation, and the concept of impact is different for the two areas.

## 2.4  Social Bots and Scholarly Articles

Leidl (2019) tried to find a correlation between the broader dissemination of research papers by bots and citations of those papers but could not find any conclusive relation between those. However, Ortega (2017) showed that the journals with their own Twitter accounts get more tweets and citations than those without.

Haustein et al. (2016) reported that automated Twitter accounts create a significant amount of tweets to scientific articles. However, the behavior of platform bots that automatically tweet links to new submissions in a particular arXiv category is the most common type of bots and behaves differently from other social bots. The authors underscore the importance of additional research into this area of alternate metrics and distinguishing the types of bots when evaluating the impact of tweets on scholarly articles. The findings from Ye and Na (2020) also report the differences between tweets from the bot and non-bot accounts. The authors found a prevalence of bot activity while analyzing Tweets containing scholarly articles related to COVID-19, and they encountered some differences with non-bot account activities. They noticed that bots use more hashtags and usually share open-access articles.

In the pursuit of detecting bots in scholarly tweets, Aljohani et al. (2020) proposed deep learning-based bot detection techniques. The authors analyzed the Altmetric Twitter Social Network and used a deep graph neural network to identify bots. They achieved 70% accuracy (F1 score of 0.67) at 200 epochs.

Arroyo-Machado et al. (2023) performed a large-scale analysis of scientific articles to identify the presence and impact of bot activity. They found that the presence of bot activity varies in different scientific disciplines. While the amount of bot activity is not significant enough to disregard Twitter, it warrants caution while using tweets as an alternate metric. The authors also used Botometer lite, which may have limited performance due to restrictions on Twitter API.

The analysis of scientific tweets by Didegah et al. (2018) found a significant presence of bot activity in certain research domains. Although the impact of dissemination of scientific articles on social media is inconclusive. After analyzing the top 10 tweeted scientific dental papers, Robinson-Garcia et al. (2017) also found that the majority of the tweets were mechanical in nature without adding any value to the conversation surrounding the papers. They suggest caution while using tweet counts as a method of evaluating the impact of scientific papers.

Even though the impact of social media activity on academic achievement is inconclusive (Didegah et al., 2018; Costas et al., 2015b; Robinson-Garcia et al., 2017), social media is still a powerful platform to bridge the communication gap between researchers and the general public. Even though the immediate engagement and feedback encourage sharing on social media (Priem and Costello, 2010; Schnitzler et al., 2016; Wadhwa et al., 2017), this can give a false perception of the public interest. Massive amounts of bot activities can also play an important role and skew the public interpretation of scientific research. This spamming by bots can shift the focus from pressing scientific research and misrepresent scientific findings. Given the bots' capability of manipulating social discussion, a better understanding of why and which research areas the bots target is essential to prevent that.

## 3  Methods

### 3.1  Data Collection

In this study, we used data from altmetric.com (Altmetrics, 2015), Twitter API (Twitter, 2020), and Botometer (on Social Media, 2016). Altmetric provides alternative metrics based on online activity for academic articles, including social media reach, blog posts, news mentions, videos, and forums. For our analysis, we leveraged a comprehensive dataset obtained from Altmetric's 2018 data snapshot. This $27 GB$ dataset encompassed all academic articles within their database until $July 4, 2018$, and included rich metadata associated with each article, such as source, DOI, publisher, and more. Additionally, it provided social media engagement metrics for each article, including the number of shares on various platforms and corresponding post identifiers. We used the data specific to Twitter from the Altmetric dataset. We used the Hydrator (Documenting the Now, 2020) tool, which leverages Twitter API, to collect information about the user who posted the article on Twitter. Finally, the Botometer API gave us a bot score for each Twitter handle. Due to the rate limits of the platforms, we developed a custom script in Python to collect data over time and populate our records. The complete data collection process took around ten weeks.

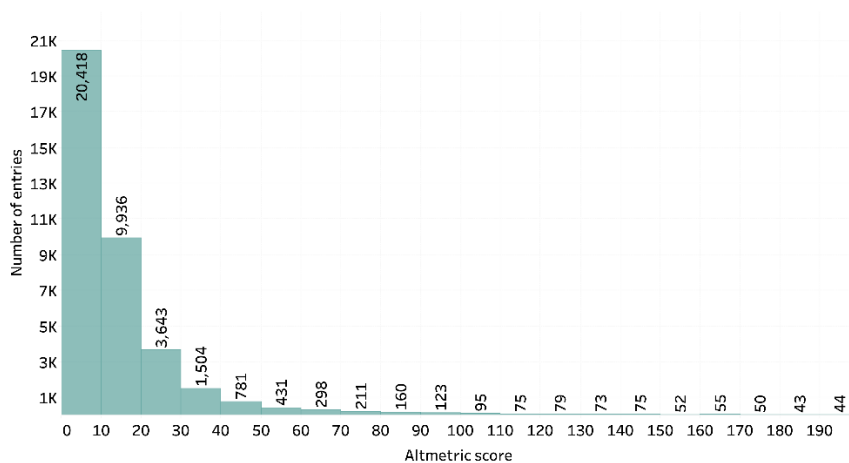

Figure 1: Altmetric score distribution (figure by authors).

The dataset from Altmetric contained 10,120,162 tweets. After removing the unavailable, suspended, and deleted accounts, we found 182,277 unique Twitter accounts that tweeted 1,398,007 academic articles. We collected information about all these Twitter accounts using the mentioned method above. Finally, we used the scores received from the Botometer API to assign a bot score to each user account.

### 3.2   Data Processing

The Botometer API provides a bot score for any given Twitter account based on several metrics such as their current activity, the historical tweets of the user, and profile information. We get a score on a scale of 0 to 5 for the metrics displayed in Table 1.

Existing research suggests that bots often tweet URLs (Chen et al., 2017; Davoudi et al., 2020). Since scholarly tweets from Altmetric usually contain the URL to the publication or journal, which can potentially bias the predictions of Botometer, we cannot readily accept the bot prediction from the Botometer. This encouraged us to calculate the bot score from individual metrics and set a threshold of bot classification manually.

Table 1: Metrics of scores provided by Botometer (table by authors).

| Metric | Description |
| --- | --- |
| Content | A feature based on linguistic cues and part-of-speech tagging. |
| Language | Language based score for the tweets, especially if the content is in English. |
| Friend | A features based on the account's social contacts and the distributions of their number of followers, followees, posts, and so on. |
| Network | This features capture information diffusion patterns based on retweets, mentions, and hashtag co-occurrence, and so on. |
| Sentiment | A feature using general-purpose and Twitter specific sentiment analysis algorithms to provide a sentiment score. |
| Temporal | A feature representing the timing patterns of content generation and consumption. |
| Universal | A feature representing an overall score for the user based on their activity on Twitter. |
| User | This feature analyzes Twitter meta-data related to an account, including language, geographic locations, and account creation time to provide a score. |

We calculated a bot score between 0 and 40 by summing up the scores for each Twitter account in our dataset. Then, we aggregated the dataset into groups of unique articles based on the Altmetric ID and created a new "overall score" feature based on the median bot score of all the tweets on an article. To mitigate biases from outliers or zero scores, we selected to use median over mean or geometric mean. This aggregation resulted in around 1.4 million records containing an overall bot activity score for each scholarly article. After this process, the features we got are listed in Table 7.

To optimize model performance, we addressed feature redundancy by analyzing similar features, like "Scopus" and "subjects," or "journal" and "publisher," and chose features covering a broader range of information. For example, "Scopus" for an article can be "Medicine," whereas the "subjects" for the paper can be "Orthopedic." We choose to keep the "Scopus" feature since it covers a broader area of research. We have also selected the primary research area from the "Scopus" feature if multiple disciplines were present. Similarly, all the Twitter user features, like follower count, user description, etc., are taken into account by Botometer. So, we removed the unnecessary features before building the model.

Also, to prevent ambiguity with the established citation database Scopus (Elsevier, 2023), we will henceforth refer to the "Scopus" feature within our analysis as "research discipline" or simply "discipline."

Table 2: List of features in the Twitter dataset that are used for training (table by authors).

| Feature | Description | Unique values |
| --- | --- | --- |
| Research discipline | Research area of the paper | 27 |
| Journal | Journal that published the paper | 3,748 |
| Research type | Whether the paper is an article, book or news | 5 |
| Publisher | Publishing company of the research | 1,821 |
| Altmetric score | Altmetric attention score for the paper based on all online activity | 39,055 |
| Author location | Geographic location of the Twitter user | 216 |
| Is Spammed | Binary feature to determine whether an article has high bot activity | 2 |

After analyzing the overall bot score feature, we noticed that the maximum score was 38.5, and the minimum was 0, with 75% of the users below the score of 16. To find an optimal threshold for isolating bots from human users, we considered the threshold levels between 15 and 25. Two Expert annotators, both graduate students, individually evaluated 30 random Twitter user accounts from each of these threshold levels for bot-like behavior, evaluating their tweet frequency, quality, content, and profile information. They unanimously agreed that the threshold of 20 is optimal for isolating bot-like users from regular users. Based on this bot score analysis, we created a new binary feature called "is Spammed," consisting of a "True" value when the overall bot score is above 20. This new feature indicates the level of bot activity towards that article. This threshold of 20 gives us 201,679 articles flagged as having higher bot activity and 1,196,328 as having low bot activity.

We noticed that the Alemetric score is vastly imbalanced, ranging from 0.25 to 8,268.56, with a mean at 114.61 and standard deviation of 326.36, as displayed in Figure 1. To avoid biasing the model by these highly skewed values, we normalized the value between the 0 and 1 ranges using a linear transformation, preserving the relationship between original data values.

At the end of these steps, our final dataset, the *Twitter dataset*, had seven features, as listed in Table 2.

### 3.3    Data Analysis

A correlation matrix, as displayed in Figure 2, of the features in the *Twitter Dataset* gives a clear picture of the relation among them. None of the features are highly correlated to our dependent variable (*Is Spammed*). There is some correlation between *Scopus*, *Journal*, and *Publisher*, but within an acceptable level for training a machine-learning model.

While we analyzed the dataset based on research areas, we noticed that the median bot score is highest in "Immunology and Microbiology," followed by "Energy" and "Pharmacy, Toxicity, and Pharmaceutics," as displayed in Figure 3.

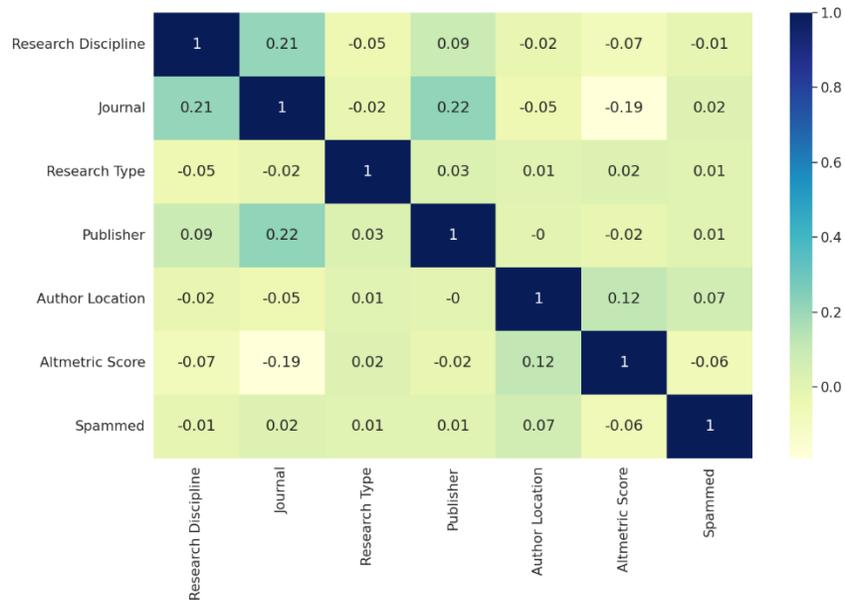

Figure 2: Correlation matrix of the dataset (figure by authors).

From our analysis, we noticed that papers from health and human science disciplines are more frequent and consequently have relatively more bot activity, as depicted in Figure 3 and Figure 4. Existing research from Didegah et al. (2018) also supports this observation. To understand if the difference in bot activity is significant, we isolated the articles for those disciplines by filtering the *Twitter dataset* and considering only the disciplines listed below.

- Biochemistry
- Genetics and Molecular Biology
- Medicine
- Life Sciences
- Health Sciences
- Psychology
- Dentistry
- Health Professions
- Nursing
- Pharmacology, Toxicology, and Pharmaceutics
- Immunology and Microbiology
- Neuroscience

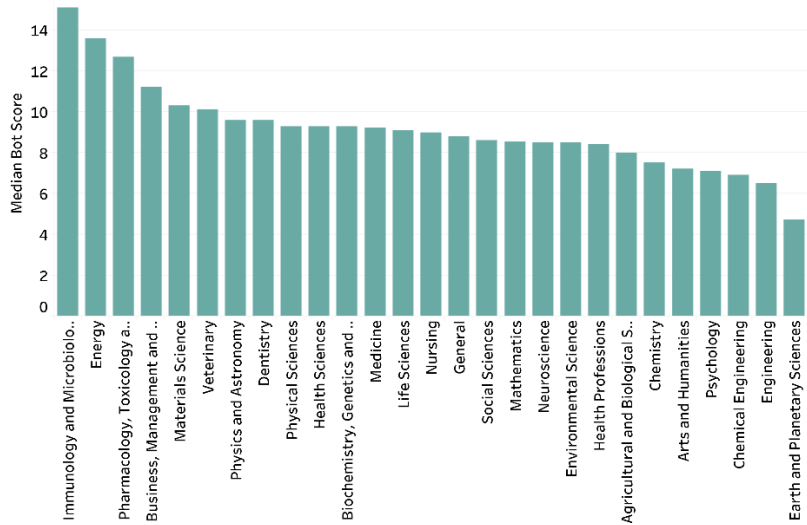

Figure 3: Median bot scores by discipline (figure by authors).

This filtering resulted in 1.17 million records out of the original 1.4 million. We will refer to this filtered data as the "*Health dataset*."

## 3.4 Building the Models

With the bot score threshold of 20, we had around 15% of the data marked as having high bot activity. We upsampled these minority class entries using Scikit-Learn's resample method to create a balanced dataset. Then, we built classifier models to predict the probability of bot activity of an article.

We build three machine-learning models for this purpose. The first model was based on K-nearest neighbors (KNN), which is a classifier model that uses proximity to make predictions (IBM, 2024a). The second model was based on support vector machine (SVM), which finds a hyperplane to maximize the distance between each class (IBM, 2024b). The third model was based on logistic regression (LR), where the statistical probability of the dependent variable is calculated based on a weighted relationship between independent variables and dependent variables (IBM, 2024c). We used the F1 score to measure the performance of the models, which represents the harmonic mean of precision and recall. An F1 score of 1 represents the perfect model, whereas a score of zero represents the opposite.

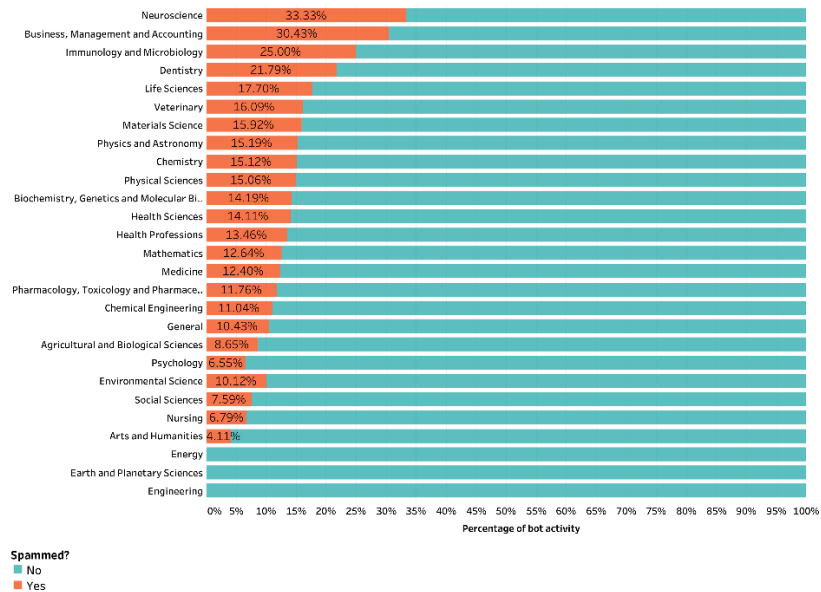

Figure 4: Percentage of bot activity by discipline (figure by authors).

With a 70 – 30 train-test split, we built these three different models using KNN, SVM, and LR. The performance of these models is listed in Table 3. The SVM performed worst with a 0.51 f1-score, followed by KNN ($n$ = 34 neighbors) with an f1-score of 0.65 and Logistic regression with an f1-score of 0.70. We tried different neighbor sizes for KNN, and the neighbor size 34 produced the best result. Table 4 and Figure 5 show the classification matrix and the ROC curve for the LR model, respectively.

Table 3: Performance of the machine-learning models (table by authors).

| Model | F1 score |
|---|---|
| KNN (n=34) | 0.65 |
| **LR** | **0.70** |
| SVM | 0.51 |

Table 4: Classification report for Logistic Regression on the *Twitter dataset* (table by authors).

|  | precision | recall | F1-score |
|---|---|---|---|
| False | 0.69 | 0.72 | 0.70 |
| True | 0.71 | 0.67 | 0.69 |
|  |  |  |  |
| accuracy |  |  | 0.70 |
| macro avg | 0.70 | 0.70 | 0.70 |
| weighted avg | 0.70 | 0.70 | 0.70 |

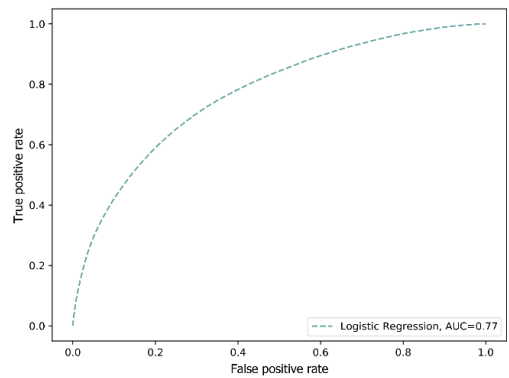

Figure 5: ROC curve for Logistic Regression on the *Twitter dataset* (figure by authors).

Then, we used the same LR model on the *Health dataset* and had an f1-score of 0.70. Table 5 and Figure 6 show this model's classification report and ROC curve.

The performance is almost identical on the *Health dataset* and the *Twitter dataset*. We expected this similarity in performance since the *Health dataset* is the majority subset of the *Twitter dataset*. We also trained a LR model with only the entries that do not belong to the Health dataset, and the model performance was the same with an f1-score of 0.70.

Table 5: Classification report for Logistic Regression on the *Health dataset* (table by authors).

|  | precision | recall | F1-score |
|---:|---:|---:|---:|
| False | 0.69 | 0.72 | 0.71 |
| True | 0.71 | 0.67 | 0.69 |
|  |  |  |  |
| accuracy |  |  | 0.70 |
| macro avg | 0.70 | 0.70 | 0.70 |
| weighted avg | 0.70 | 0.70 | 0.70 |

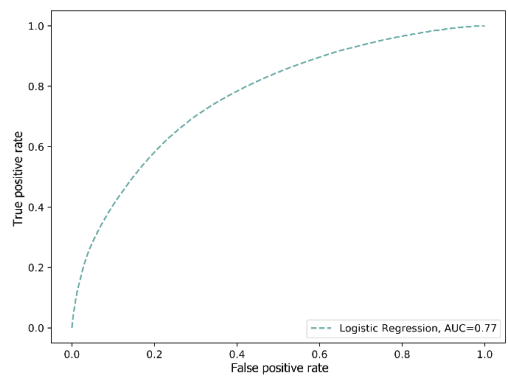

Figure 6: ROC curve for Logistic Regression on the *Health dataset* (figure by authors).

While looking at the feature importance from the trained LR model as depicted in Figure 7, we found that *research type* and *research discipline* were the most important features for the predictor, with a median coefficient value of 0.06 each, followed by *journal*, and *author location*, with a median coefficient value of −0.06. We also noticed a relatively higher negative coefficient value for *Altmetric score*, a median coefficient value of −0.20, indicating less impact of bot activity on articles with high Altmetric score.

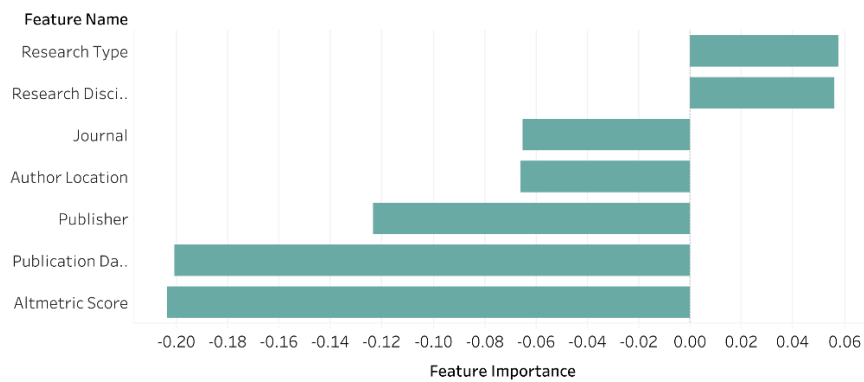

Figure 7: Feature importance of the trained LR model (figure by authors).

## 3.5 Statistical Analysis

We observed from basic plots in Figure 3 and Figure 4 that health & human science-related articles are more likely to have bot activity. We performed a hypothesis test to determine whether the higher level of bot activity is statistically significant. For this purpose, we considered the following hypothesis:

**Null hypothesis:**

Bot activity in health and human science is not significantly higher compared to other research disciplines.

**Alternate hypothesis:**

Health and human science articles have more bot activity than the other research disciplines.

If we plot the ratio of articles with high bot activity, as displayed in Figure 8, we can see that the health and human science discipline has slightly higher bot activity.

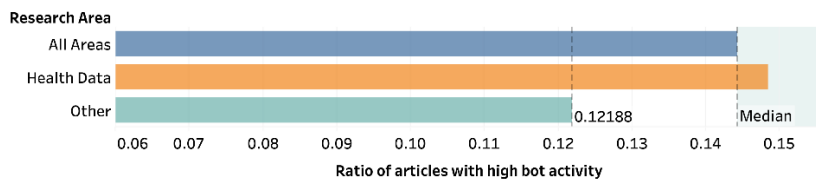

Figure 8: Comparison of the ratio of bot activity on all research articles, health & human science areas, and other areas in the original dataset (figure by authors).

We performed a z-test, a two-tailed hypothesis test, to validate our hypothesis. We considered a confidence level of 99% required to reject the null hypothesis. So, the p-value must be smaller than 0.01. Table 6 shows all the data points we had for the analysis.

Table 6: Data Points for statistical analysis (table by authors).

| Datapoint | | Value |
|---|---|---|
| All research areas | Number of articles | 1,298,007 |
| | Articles with high bot activity | 201,679 (14.43%) |
| Health & human science areas | Number of articles | 1,178,085 |
| | Articles with high bot activity | 174,876 (14.84%) |
| Other areas | Number of articles | 219,922 |
| | Articles with high bot activity | 26,803 (12.19%) |

Once we performed the z-test, we got a p-value less than 0.001 and a z-score of 32.5; 32.5 standard deviations to the right of the center. Based on this test, we could reject the null hypothesis and conclude that health and human sciences articles have significantly more bot activity than other research areas.

## 4 Discussion

To address our first research question (RQ: 1), the proposed model only utilizes Altmetric metadata to reliably predict the possibility of bot activity on Twitter in the dissemination of any scholarly article. Compared to deep learning approaches (Aljohani et al., 2020), our model achieves superior performance while requiring fewer computational resources, ensuring efficiency and accessibility.

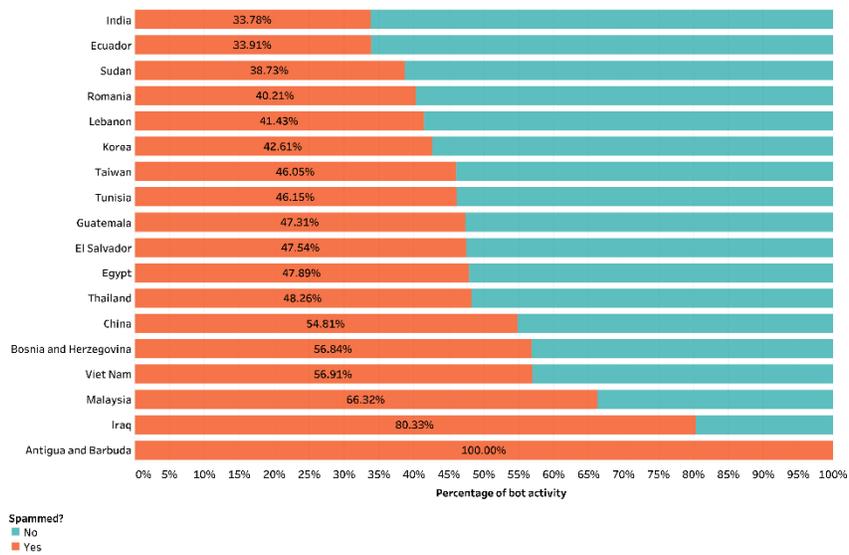

Figure 9: Countries with a high percentage (> 30%) of bot activity (figure by authors).

This study gives us a critical perspective on disseminating scholarly articles on social media. While there is an increasing distrust toward science in society for different reasons (Hamilton and Safford, 2021; Contessa, 2022; Bromme and Hendriks, 2022), the model we developed can validate the popularity of any given scholarly article on social media and can work as a baseline for future research in this direction. This model does not argue the authenticity or validity of the actual work, nor does it claim the bot activity as either good or bad. Instead, this model provides a metric to determine whether the dissemination of an academic article has the possibility to be impacted by bot activity. This prediction can be helpful for both researchers and policymakers to understand the public acceptance of any given scientific publication.

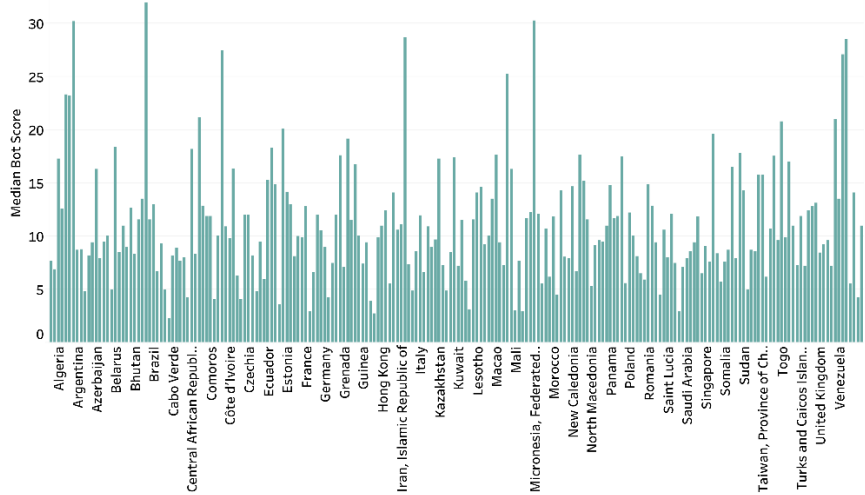

Figure 10: Median bot scores by country (figure by authors).

We noticed some distinctive trends in the bot activity while analyzing the dataset. By grouping our data based on user locations, we observed a high number of bot activities in the USA, UK, France, India, and many European countries. This higher frequency of bot activity is expected because of the higher adoption rate of Twitter in these countries. However, we noticed a different trend when looking at the percentage of bot activity in the tweets, as displayed in Figure 9. During our analysis, we recognized that bot activity in academic tweets is significantly higher in some developing countries like China (54.81%), Egypt (47.89%), India (33.78%), Iraq (80.33%), Malaysia (66.32%), Taiwan (46.05%), Thailand (48.26%), Vietnam (56.91%), and so on. We intend to do a more in-depth study to understand the possible reason behind this high bot activity ratio in certain regions.

We also noticed that the median bot score is relatively high in the countries with higher bot activity ratios while others are moderately below our threshold of 20, as displayed in Figure 10.

We extrapolated that research type and research discipline were the most important predictors for the probability of bot activity. Surprisingly, the Altmetric score presented a negative coefficient to the prediction, indicating that articles with higher Altmetric scores have less impact from bot activity.

We also identified that the high level of bot activity in health and human science-related disciplines is statistically significant, hence answering our second research question (RQ: 2). This finding can give an important perspective to the research community in identifying and addressing bot activity in scientific discussions on social media.

The proposed model from this study stays relevant even after potential limitations imposed by API restrictions from social media platforms. By leveraging solely the metadata obtainable from Altmetric, the model maintains its functionality regardless of future API changes. Additionally, the model serves as a benchmark for future research in bot activity in academic publications and similar research on social media platforms other than Twitter.

## 5 Limitations and Future Works

We considered the features available from the Altmetric dataset. It can be exciting research to extract additional features about the authors of the article, the location of the publication, international collaboration, and other demographic features of the authors to see the relation of these features with bot activity.

Our study concluded that the bots targeted the health and human science disciplines at a higher rate. Although we could not identify a definitive reason for that, we hypothesize that engaging the public with health and human science is much easier than in other advanced scientific areas. People care about diseases, vaccination, and the healthcare systems, attracting more bots and spamming in this area. Further research in this direction can uncover meaningful insights.

Besides the overall bot activity, our current work only evaluated bot activity in the health and human science area. Further analysis to compare bot activity between different scientific areas can enrich our understanding of bot activity in academia.

## 6 Conclusion

In this study, we examined how Altmetrics features combined with Twitter users' information can predict if a scholarly paper will attract bots on Twitter. Previous work does not explicitly find a link between Twitter bots and academic articles. We have taken several features from Altmetrics, Twitter, and Botometer and fed them into a logistic regression model. We achieved 70% accuracy in predicting bot activity on any scholarly article by Twitter bots. We have also analyzed and concluded that there is significantly higher bot activity in the health and human science disciplines than in other research areas.

Since public interest in scientific findings can shape the decisions of policymakers, it is essential to identify the possibility of bot activity in the dissemination of any given scholarly article. Our work proposes a machine-learning model to interpret the public interest in any scientific article by predicting the possibility of bot activity in the article's dissemination. We published and made the models and data freely available for the research community, providing a benchmark and guideline for future works in this direction.

## Supplementary Information

The pre-trained model and the dataset are available to download from the links provided in the appendix.

## Acknowledgments

We appreciate Venkata Devesh Reddy Seethi (ORCID: 0000-0002-7400-7518) for his help with the Altmetric data.

## Declarations

### Funding

This work is supported in part by NSF Grant No. 2022443 and by the Office of the Vice President for Research at the University of South Carolina.

### Competing Interests

The authors have no competing interests to declare that are relevant to the content of this article.

# Appendix A     Machine Learning Model

## A.1     Spam Prediction Model

A pre-trained logistic regression model to predict the possibility of an article being spammed or not.

The model is available at: https://doi.org/10.5281/zenodo.7823566

# Appendix B     Datasets

## B.1     Labeled Twitter Dataset

The dataset contains the Altmetric IDs of the articles along with the overall bot score, Scopus, and possible spammed classification.

The dataset is available at: https://doi.org/10.5281/zenodo.7823566

## B.2     Original Dataset Features

The features available in the original dataset before processing are listed in Table 7.

Table 7: Features available in the dataset before cleanup (table by authors).

| Feature | Description |
| --- | --- |
| Altmetric ID | Unique ID for each entry |
| Scopus (Research discipline) | Research area of the paper |
| Twitter poster types | An indication of the tweet author (e.g., researcher, science communicator, public, etc.) |
| Paper pubdate | Publication date of the paper |
| First seen on | First time seen on social media |
| Last mentioned on | Last time seen on social media |
| Subjects | Subjects covered in the paper (sub-groups of "scopus" above) |
| Selected quotes | Text quoted in the tweet along with the paper's link |
| Funders | Funders of the paper |
| Twitter unique users count | Number of unique Twitter users sharing the paper |
| Twitter posts count | Number of tweets sharing the paper |
| Journal | Journal that published the paper |
| Research type | Whether the paper is an article, book, or news |
| Publisher | Publishing company of the research |
| Altmetric score | Altmetric attention score for the paper based on all online activity |
| Authors | List of authors of the paper |
| Tweet ID | Unique tweet ID from the Twitter API |
| Twitter desc | Profile description of the Twitter user |
| Twitter ID | Unique user ID of the Twitter user |
| Twitter author followers | Number of followers of the Twitter user |
| Twitter author name | Name of the Twitter user |
| Author loc | Geographic location of the Twitter user |
| Tweet posted on | Date of the tweet |
| Retweeters | Twitter users who retweet the original tweet |
| Author ID | Twitter handle of the user |
| Overall score | Overall Botometer score for the article |